\documentclass[fleqn,11pt,twoside]{article}
\usepackage{amsmath}
\usepackage{amssymb}
\usepackage{espcrc1}
\usepackage{graphicx}

\title{QCD calculations of thermal photon and dilepton production}

\author{Fran\c cois Gelis\address{Laboratoire de Physique Th\'eorique, \\ 
        Universit\'e Paris XI, B\^atiment 210, \\ 
        91405, Orsay Cedex, France}%
        \thanks{Talk given at Quark-Matter 2002, July 18-24 2002, Nantes, France}}
       
\begin{document}

\maketitle

\begin{abstract}
  In this talk, I review new developments of QCD calculations of
  photon and dilepton production rates in a Quark-Gluon plasma. All
  the rates are now known up to ${\cal O}(\alpha_s)$ both for
  photons and dileptons, thanks to the resummation of multiple
  scatterings. For dileptons, a direct numerical calculation on the
  lattice attempted recently will also be discussed.
\end{abstract}

\section{INTRODUCTION}
Electromagnetic probes (photons or lepton pairs) have production rates
that are very sensitive to the temperature of the medium in which they
are produced (the larger the temperature, the larger the rate) and are
therefore mostly produced during the very early stages of a heavy ion
collision. In addition to being produced early, they are weakly
coupled to nuclear matter and have therefore a mean free path which is
large compared to the typical size of the system produced in a
collision. This enables them to escape from the system without any
reinteraction. These two properties combined make electromagnetic
signals very good probes of the state of the system very early after
the initial impact.

\begin{figure}[htbp]
\begin{flushleft}
\resizebox*{!}{3.8cm}{\includegraphics{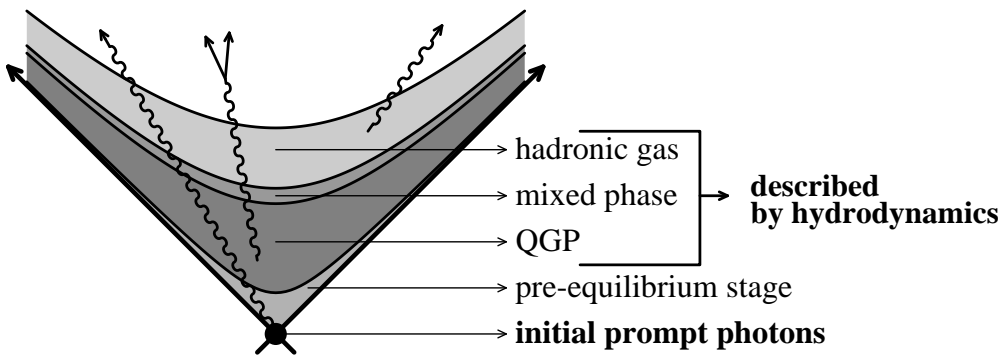}}
\end{flushleft}
\caption{\label{fig:sketch} The various stages of a heavy ion collision.}
\end{figure}

Roughly speaking, a heavy ion collision can be divided into several
distinct stages, as illustrated on figure \ref{fig:sketch}.
Some photons are produced in the initial partonic collisions. These
prompt photons can be calculated using zero temperature perturbative
QCD, and they populate the high energy part of the spectrum.  Then
comes a pre-equilibrium phase, which is usually thought to be very
short and very poor in quarks and anti-quarks, so that photon
production in this phase is neglected. In scenarios with a quark-gluon
plasma, this is followed by the plasma phase, which produces photons
at a rate calculable using equilibrium thermal field theory (TFT).
Indeed, a local equilibrium is usually assumed for the plasma phase,
and the local rate calculated using TFT is then folded in an
hydrodynamical evolution code \cite{Sriva1,SrivaS1,HuoviRR1,AlamSRHS1}
in order to perform the integration over space-time. After the
confinement phase transition, the system becomes a hadronic gas (there
can be a mixed phase if the phase transition is first order), for
which TFT (now with hadronic degrees of freedom) can also be used in
order to compute the photon rate. Finally the system freezes out and
the only photons that are produced afterwards are decay products of
hadrons. The photon yield observed in detectors is the sum of all
these contributions. In the rest of this talk, I focus on photons
produced in the QGP phase, and on the TFT techniques used in order to
compute their rate.

In order to compute the photon production rate in a QGP, one can
proceed as follows: make the list of the processes contributing up to
a given order, add up the corresponding amplitudes, square this sum
and integrate out all the particles but the photon, properly weighted
by the appropriate statistical distribution.
However, this method becomes rapidly cumbersome since one must track
by hand the statistical factors, and the interference terms. TFT
provides a convenient alternative to this approach, which has the
advantage of taking care automatically of the statistical factors and
interferences. In this approach, the photon production rate is
expressed in terms of the imaginary part of the retarded polarization
tensor \cite{Weldo3,GaleK1}:
\begin{equation}
    { {\omega\frac{dN_\gamma}{dt dV d^3{\boldsymbol q}}}}=
 \frac{1}{(2\pi)^3} \frac{1}{e^{\omega/T}-1}
\;{ {\rm Im}\,\Pi_{\rm ret}{ {}^\mu{}_\mu}(\omega,{\boldsymbol q})}\; .
\label{eq:rate-1}
\end{equation}
The calculation of this retarded imaginary part can be simplified by
the use of cutting rules \cite{Gelis3}, and it is the sum over all the
possible cuts that takes care of the interference terms. Note that a
similar formula exists for lepton pairs, for which one needs to
compute the polarization tensor of a massive photon (i.e. $Q^2\equiv
\omega^2-{\boldsymbol q}^2>0$).

\section{HISTORY}
\begin{figure}[htbp]
\begin{flushleft}
\resizebox*{!}{0.5cm}{\includegraphics{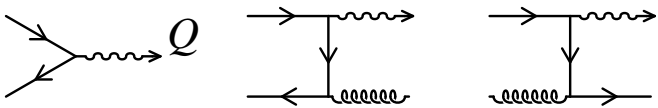}}
\end{flushleft}
\caption{\label{fig:processes-1} Real processes contributing to 
  photon and dilepton production up to ${\cal O}(\alpha_s)$.
  Virtual corrections to the first process also contribute at this
  order.}
\end{figure}

The calculation of thermal photon and dilepton rates has a long and
tortuous history. The dilepton rate due to the Drell-Yan process (see
the diagram on the left of figure \ref{fig:processes-1}) was evaluated
in a QGP in \cite{McLerT1}.
Corrections of order ${\cal O}(\alpha_s)$ were considered shortly
afterwards, and two problems became apparent. If one tries to
calculate the dilepton rate ($Q^2>0$) in a plasma of massless quarks
and gluons, each individual cut contributing to eq.~(\ref{eq:rate-1})
contains a mass singularity, and it is only after a careful summation
of all the real and virtual corrections that one gets a finite result
\cite{BaierPS2,AltheAB1,AltheR1}.  This is nothing but a manifestation
of a general property noted in \cite{LeeN1}. If one then tries to take
the limit of real photons ($Q^2\to 0^+$) in the above formula, a new
singularity appears since there are some terms that behave like ${\rm
  Im}\,\Pi_{\rm ret}(\omega,{\boldsymbol q}) \propto \alpha\alpha_s
\ln(\omega T/Q^2)$ at small $Q^2$.

The latter problem was resolved by the resummation of Hard Thermal
Loops (HTL). HTL are one-loop leading thermal corrections that have
the same order of magnitude as their bare counterpart when their
external momenta are soft (i.e. of ${\cal O}(\sqrt{\alpha_s}T)$).
They were known for quite some time in the case of 2-point functions
\cite{Klimo2,Weldo2}, but a systematic gauge invariant resummation was
only proposed in 1990 by \cite{BraatP1,FrenkT1}. In the present case,
the logarithmic singularity at $Q^2\to 0$ is due to the exchange of a
soft massless quark. This singularity is screened by the resummation
of the HTL correction to the quark propagator, which gives the quark a
thermal mass $m_{\rm q}$ of order $m_{\rm q}\sim \sqrt{\alpha_s}T$.
Taking into account this thermal correction to the quark propagator
solves the problem and leads to a finite photon polarization tensor
\cite{KapusLS1,BaierNNR1}. For hard photons, it reads:
\begin{equation}
{\rm Im}\,\Pi_{\rm ret}{}^\mu{}_\mu(\omega,{\boldsymbol q})=4\pi 
{ \frac{5\alpha\alpha_s}{9}}T^2
\left[\ln\left(\frac{\omega T}{m_{\rm q}^2}\right)
-\frac{1}{2}-\gamma_{_{E}}+\frac{7}{3}\ln(2)+\frac{\zeta^\prime(2)}{\zeta(2)}
\right]\; .
\end{equation}
Note that throughout this talk the mass $m_{\rm q}$ is defined to be
the asymptotic quark thermal mass, i.e. $m_{\rm q}^2=\pi\alpha_s C_f
T^2$ with $C_f\equiv(N_c^2-1)/2N_c$.  The numerical factor $5/9$ is
the sum of the quark electric charges squared for $2$ flavors (u and
d); for $3$ flavors (u, d and s), this factor should be replaced by
$6/9$.  Regarding the infrared problem, one can see that $Q^2$ is
replaced by $m_{\rm q}^2$ in the logarithm as soon as $Q^2$ becomes
small compared to $m_{\rm q}^2$.

\begin{figure}[htbp]
\begin{flushleft}
\resizebox*{!}{1cm}{\includegraphics{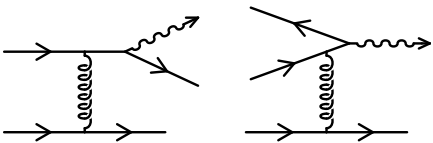}}
\end{flushleft}
\caption{\label{fig:processes-2} Two processes that are 
  promoted to ${\cal O}(\alpha_s)$ by collinear singularities.}
\end{figure}

This was thought to be the final answer for the photon and dilepton
rates at ${\cal O}(\alpha_s)$, until it became clear that some
formally higher order processes are in fact strongly enhanced by
collinear singularities. This was first realized for soft photon
production by quark bremsstrahlung \cite{AurenGKP1,AurenGKP2} (left
diagram of figure \ref{fig:processes-2}).  The diagram on the right of
figure \ref{fig:processes-2} shares the same property, but contributes
significantly only to hard photon production \cite{AurenGKZ1}, due to
phase-space suppression in the case of soft photons. Note that a naive
power counting would indicate that these two diagrams contribute to
${\cal O}(\alpha_s^2)$.  A common property of these two diagrams is
that they have an off-shell quark next to the vertex where the photon
is emitted, and that the virtuality of this quark can become very
small if the photon is emitted forward.
Again, it is the quark thermal mass $m_{\rm q}$ that prevents these
diagrams from being truly singular. However, contrary to the ${\cal
  O}(\alpha_s)$ diagrams, the singularity is linear instead of
logarithmic, and brings a factor $T^2/m_{\rm q}^2$. Combined with the
$\alpha_s^2$ that comes from the vertices, these diagrams turn
out to be also of order ${\cal O}(\alpha_s)$. In \cite{AurenGKP2}
and \cite{AurenGKZ1}, their contribution was derived semi-analytically,
with a prefactor that was evaluated numerically. \cite{Mohan2} and
\cite{SteffT1} independently pointed out an erroneous factor $4$ in
this numerical prefactor. Finally, it was realized that it can be
calculated in closed form \cite{AurenGZ4,AurenGZ3}. For 3 colors and 2
light quark flavors, the ${\cal O}(\alpha_s)$ contribution of
these two diagrams is exactly:
\begin{equation}
{\rm Im}\,\Pi_{\rm ret}{}^\mu{}_\mu(\omega,{\boldsymbol q})=
\frac{32}{3\pi} \,
\frac{5\alpha\alpha_s}{9}\left[\pi^2 \frac{T^3}{\omega}+\omega T\right]\; .
\end{equation}
In this formula, the term in $1/\omega$ dominates for soft photons and
comes from the brem\-sstra\-hlung diagram, while the term in $\omega$
comes from the second diagram. Because of this term in $\omega$, this
process turns out to dominate the rate of very hard photons ($\omega
\gg T$). This was confirmed by more realistic evaluations that
included this local rate into an hydrodynamical evolution code, and
there is some speculation that these hard thermal photons could be
part of the excess of direct photons observed by the WA98 experiment
at SPS \cite{Sriva1,SrivaS1,HuoviRR1,chaudhuri,chaudhuri1,Aggara1}
(see also \cite{panel}).  Note that it is only by accident if this
result is so simple. For 3 colors and 3 light flavors, the same
quantity can still be calculated in closed form (note that the energy
dependence is the same), but the prefactor is much more involved:
 \begin{eqnarray}
 {\rm Im}\,\Pi_{\rm ret}{}^\mu{}_\mu(\omega,{\boldsymbol q})&=&
 \frac{32}{3\pi}
 \left[
 1+\frac{5\pi^2}{36}+{\ln\left(\frac{\sqrt{2}}{3}\right)}
 -\frac{55}{12}\ln^2(2)+\frac{10}{3}\ln(2)\ln(3)\right.\nonumber\\
 &&\quad\qquad-\left.
 \frac{5}{3}{\rm Li}_2\left(\frac{3}{4}\right)
 -\frac{5}{3}{\rm Li}_2\left(-\frac{1}{2}\right)
 \right]\,\frac{6\alpha\alpha_s}{9}\left[\pi^2 \frac{T^3}{\omega}+\omega T\right]\; .
 \end{eqnarray}
 It is also worth mentioning at this point that the purely numerical
 prefactor is a function of the ratio of the quark thermal mass
 $m_{\rm q}$ to the gluon Debye mass $m_{\rm debye}$ (the Debye mass
 quantifies the effective range of strong interactions when they are
 screened by medium effects, and it is also of ${\cal
   O}(\sqrt{\alpha_s}T)$).  In the HTL framework, this ratio is a
 constant independent of the coupling and temperature, that depends
 only on the number of colors and flavors; for $3$ colors and $N_f$
 flavors, this ratio is $m_{\rm q}/m_{\rm debye}=\sqrt{2/(6+N_f)}$.

\section{LPM EFFECT}
Given the enhancement in the diagrams of figure \ref{fig:processes-2},
one may wonder if there are formally higher order diagrams that also
end up contributing to the same order ${\cal O}(\alpha_s)$. This was
partly answered in \cite{AurenGZ2} where the resummation of a
collisional width $\Gamma\sim \alpha_s T\ln(1/\alpha_s)$ on the quarks
in the calculation of the diagrams of figure \ref{fig:processes-2}
showed some sensitivity to this parameter at leading order, thereby
indicating that an infinite series of diagrams must be resummed in
order to fully determine the ${\cal O}(\alpha_s)$ photon rate.

In order to explain the issue in more physical terms, it is convenient
to define the concept of {\sl photon formation time}. Let me assume
that a virtual quark of momentum $R\equiv P+Q$ splits into an on-shell
quark of momentum $P$ and a photon of momentum $Q$. The photon
formation time can be identified with the lifetime of the virtual
quark, which is itself related to its virtuality by the uncertainty
principle. For a small $Q^2$, a simple calculation gives:
\begin{equation}
t_{_{F}}^{-1}\sim { \delta E}=r_0-\sqrt{{\boldsymbol r}^2+m_{\rm q}^2} 
\approx \frac{\omega}{2 p_0 r_0}
\left[{\boldsymbol p}_\perp^2+{ m_{\rm q}^2}+\frac{Q^2}{\omega^2}p_0 r_0
\right]\; ,
\label{eq:tF}
\end{equation}
where the 3-momentum of the photon defines the longitudinal axis. Note
that the collinear enhancement in the diagrams of figure
\ref{fig:processes-2}, due to the small virtuality of the quark that
emits the photon, can be rephrased by saying that it is due to a large
photon formation time. Similarly, the sensitivity to the collisional
width found in \cite{AurenGZ2} occurs if the photon formation time is
of the same order or larger than the quark mean free path between two
soft collisions (this mean free path $\lambda$ is the inverse of the
width $\Gamma$). This phenomenon is nothing but a manifestation of the
Landau Pomeranchuk Migdal (LPM) effect \cite{LandaP1,LandaP2,Migda1}.

\begin{figure}[h]
\begin{flushleft}
\resizebox*{6cm}{!}{\includegraphics{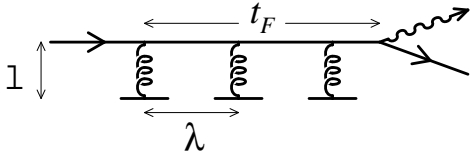}}
\end{flushleft}
\caption{\label{fig:LPM} A ladder correction to bremsstrahlung.}
\end{figure}

The precise nature of the multiple scattering diagrams that must be
resummed depends in fact on the range of the interactions in the
medium. Indeed, if the range $\ell$ of the interactions is much shorter
than the mean free path, it is easy to check that only ladder
topologies are important, in which all the successive scatterings are
independent of one another, as illustrated in figure \ref{fig:LPM}.
Indeed, the condition $\ell\ll\lambda$ suppresses diagrams with
crossed gluons. On the contrary, if there are long range interactions
in the system, for which $\ell\gtrsim \lambda $, then arbitrarily
complicated topologies can a priori contribute.
It was found in \cite{AurenGZ1} that if one considers contributions to
the photon polarization tensor topology by topology, then there can be
a sensitivity to interaction ranges as long as the magnetic scale
$1/\alpha_sT$ (for a review of the relevant scales and associated
physics in a QGP, see \cite{dietrich}), which would render the problem
practically intractable.

\begin{figure}[h]
\begin{flushleft}
\resizebox*{13cm}{!}{\includegraphics{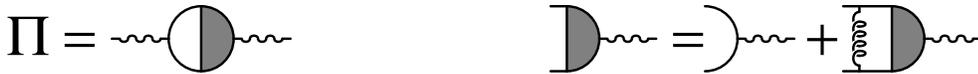}}
\end{flushleft}
\caption{\label{fig:equations}  Resummation of ladder diagrams.}
\end{figure}

\begin{figure}[htbp]
\begin{flushleft}
\resizebox*{!}{4.5cm}{\rotatebox{-90}{\includegraphics{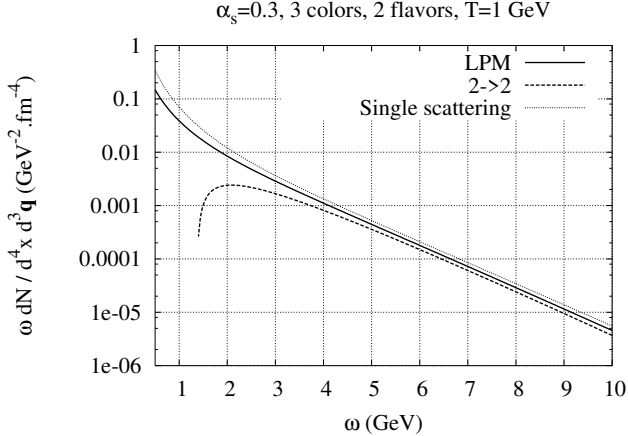}}}
\end{flushleft}
\caption{\label{fig:LPM-photon} ${\cal O}(\alpha_s)$ contributions
  to the photon production rate in a QGP. The parameters used in this
  plot are $\alpha_s=0.3$, 3 colors, 2 flavors and $T=1$~GeV.}
\end{figure}

A considerable progress was made recently in \cite{ArnolMY1}, in which
it was shown that there are infrared cancellations between diagrams of
different topologies, and that these cancellations remove any
sensitivity to the magnetic scale. Physically, this cancellation can
be interpreted as the fact that ultrasoft scatterings are not
efficient in order to induce the production of a photon. As a
consequence, only the ladder family of diagrams needs to be resummed
in order to obtain the complete leading ${\cal O}(\alpha_s)$ photon
rate. The resummation of this series of diagram can then be performed
in two steps summarized in figure \ref{fig:equations}.
The first one is a Dyson equation for the photon polarization tensor,
whose explicit form is \cite{ArnolMY1,ArnolMY2,ArnolMY3}:
\begin{eqnarray}
{\rm Im}\,\Pi_{\rm ret}{}^\mu{}_\mu(Q)\approx 
{\alpha N_c}
\int_{-\infty}^{+\infty}\!\!dp_0
\,[n_{_{F}}(r_0)-n_{_{F}}(p_0)]\;
\frac{p_0^2+r_0^2}{(p_0r_0)^2}
\,{\rm Re}\int 
\frac{d^2{\boldsymbol p}_\perp}{(2\pi)^2}\;
{\boldsymbol p}_\perp\cdot{\boldsymbol f}({\boldsymbol p}_\perp)\; ,
\label{eq:AMY}
\end{eqnarray}
with $r_0\equiv p_0+q_0$, $n_{_{F}}(p_0)\equiv 1/(\exp(p_0/T)+1)$ the
Fermi-Dirac statistical weight, and where the dimensionless function
${\boldsymbol f}({\boldsymbol p}_\perp)$ denotes the resummed vertex
between the quark line and the transverse modes of the photon (this is
represented by the shaded vertex in the above pictures). In the Dyson
equation, this function is dotted into a bare vertex, which is
proportional to ${\boldsymbol p}_\perp$.  The second equation, that
determines the value of ${\boldsymbol f}({\boldsymbol p}_\perp)$, is a
Bethe-Salpeter equation that resums all the ladder corrections
\cite{ArnolMY1,ArnolMY2,ArnolMY3}:
\begin{equation}
\frac{i}{t_{_{F}}}{\boldsymbol f}({\boldsymbol p}_\perp)
=
2{\boldsymbol p}_\perp
+4\pi \alpha_s C_f T \!\! \int
\frac{d^2{\boldsymbol l}_\perp}{(2\pi)^2} \, 
{\cal C}({\boldsymbol l}_\perp) \,
[{\boldsymbol f}({\boldsymbol p}_\perp+{\boldsymbol l}_\perp)-{\boldsymbol f}({\boldsymbol p}_\perp)]\; ,
\label{eq:integ-f}
\end{equation}
where $t_{_{F}}$ is the time defined in eq.~(\ref{eq:tF}) and where
the collision kernel has the following expression: ${\cal
  C}({\boldsymbol l}_\perp)={m^2_{\rm debye}}/{{\boldsymbol
    l}_\perp^{\ 2}} ({{\boldsymbol l}_\perp^{\ 2}+{ m^2_{\rm
      debye}}})$ \cite{AurenGZ4}.  Note that in the Dyson equation,
the quark propagators should be dressed in a way compatible with the
resummation performed for the vertex, in order to preserve the gauge
invariance. It is this dressing on the quark propagators which is
responsible for the term $-{\boldsymbol f}({\boldsymbol p}_\perp)$
under the integral in eq.~(\ref{eq:integ-f}).  From this integral
equation, it is easy to see that each extra rung in the ladder
contributes a correction of order $\alpha_sT p_0 r_0/\omega m_{\rm
  q}^2$, in which the $\alpha_s$ drops out.  Therefore, all these
corrections contribute to ${\cal O}(\alpha_s)$ to the photon rate.
Note again that the only parameters of the QGP that enter in this
equation are the quark thermal mass $m_{\rm q}$ and the Debye
screening mass $m_{\rm debye}$. This integral equation was solved
numerically in \cite{ArnolMY2}, and the results are displayed in
figure \ref{fig:LPM-photon}.
In this plot, `LPM' denotes the contribution of all the multiple
scattering diagrams, while `$2\to 2$' denotes the processes of figure
\ref{fig:processes-1}. The single scattering diagrams (figure
\ref{fig:processes-2}) are also given so that one can appreciate the
suppression due to the LPM effect.

\section{DILEPTON PRODUCTION}
Dilepton production basically suffers from the same problems, and the
solution follows the same path. Two differences are worth mentioning
here. First of all, the Drell-Yan process $q\bar{q}\to\gamma^*\to l^+
l^-$ contributes if $Q^2\ge 4 m_{\rm q}^2$. In addition, virtual
photons have a physical longitudinal mode that contributes to the rate
of lepton pairs. The Drell-Yan process has been evaluated in
\cite{McLerT1}, and the $2\to 2$ processes have been evaluated in
\cite{AltheR1}.

\begin{figure}[htbp]
\begin{flushleft}
\resizebox*{!}{4.5cm}{\rotatebox{-90}{\includegraphics{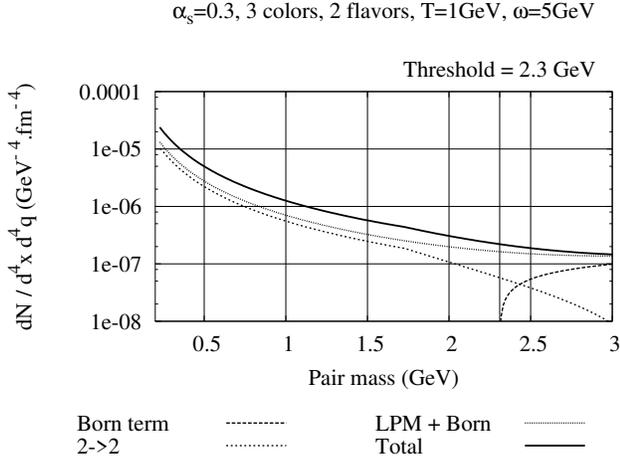}}}
\end{flushleft}
\caption{\label{fig:LPM-lepton} ${\cal O}(\alpha_s)$ contributions
  to the dilepton production rate in a QGP.}
\end{figure}

For photon invariant masses of order $Q^2 \sim \alpha_s T^2$ or
smaller, one expects also important contributions from multiple
scattering diagrams. One must now keep track of the non-zero $Q^2$,
and include also the contribution of the photon longitudinal mode.
This is easily done by performing the following substitution in
eq.~(\ref{eq:AMY}) \cite{AurenGMZ1}:
\begin{equation}
\frac{p_0^2+r_0^2}{(p_0r_0)^2}\,
{\boldsymbol p}_\perp\cdot{\boldsymbol f}({\boldsymbol p}_\perp)
\to
\frac{p_0^2+r_0^2}{(p_0r_0)^2}\,
{\boldsymbol p}_\perp\cdot{\boldsymbol f}({\boldsymbol p}_\perp)
+\frac{2}{\sqrt{|p_0r_0|}}\frac{Q^2}{q_0^2}\,g({\boldsymbol p}_\perp)\; ,
\end{equation}
where the function $g({\boldsymbol p}_\perp)$, which describes the
coupling between the quark line and the longitudinal photon, obeys an
integral equation similar to eq.~(\ref{eq:integ-f}) \cite{AurenGMZ1}:
\begin{equation}
\frac{i}{t_{_{F}}} {g}({\boldsymbol p}_\perp)
=
2\sqrt{|p_0r_0|}
+4\pi\alpha_s C_f T\int
\frac{d^2{\boldsymbol l}_\perp}{(2\pi)^2} {\cal C}({\boldsymbol l}_\perp)
[{g}({\boldsymbol p}_\perp+{\boldsymbol l}_\perp)-{g}({\boldsymbol p}_\perp)]\; .
\label{eq:integ-g}
\end{equation}
Note that the contribution of the longitudinal mode of the photon
vanishes trivially when $Q^2\to 0$, as it should. This new integral
equation can also be solved numerically, and the resulting dilepton
rate (for the same parameters as in figure \ref{fig:LPM-photon} and a
total energy of the pair set to $\omega=5~$GeV) is plotted in figure
\ref{fig:LPM-lepton}.
One can see that the multiple scattering corrections are important for
all pair masses below the threshold of the Drell-Yan process. Note
also that the threshold of the tree-level process is completely washed
out when multiple rescatterings are resummed.

\section{QUASIPARTICLE MODELS}
We have emphasized several times the fact that the only properties of
the QGP that these rates depend on are the quark thermal mass $m_{\rm
  q}$ and the Debye mass $m_{\rm debye}$. So far, these two masses
have been taken in the HTL approximation, for which the ratio of the
two masses is independent of $T$ and $\alpha_s$. However, simple
arguments indicate that this ratio cannot remain constant when the
mass $m_{\rm q}$ becomes large, which may happen at moderate
temperatures for which the coupling constant is rather large. Indeed,
the Debye screening is due to the possibility for a test charge to
polarize the medium surrounding it in order to screen its charge. This
process becomes difficult to achieve when the quasiparticles in the
medium become very heavy, and for this reason $m_{\rm debye}$ should
become very small if $m_{\rm q}$ increases. This is indeed what one
finds by calculating the Debye mass at 1-loop, with massive particles
running in the loop.

\begin{figure}[htbp]
\begin{flushleft}
\resizebox*{!}{4cm}{\includegraphics{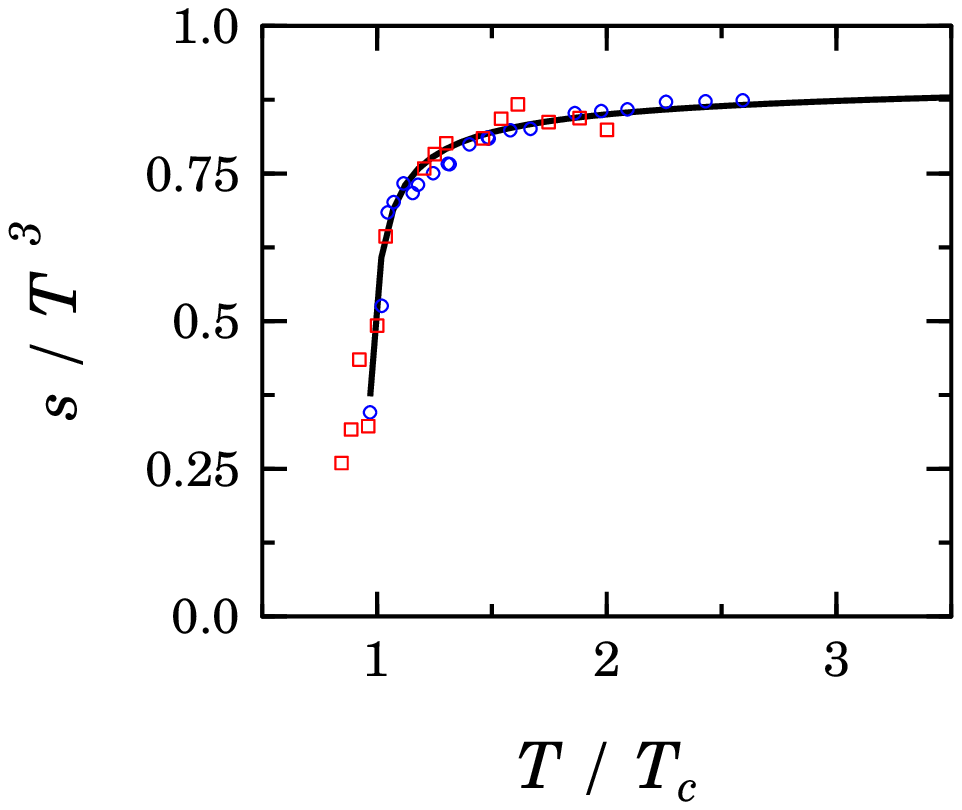}}
\hglue 3mm
\resizebox*{!}{4cm}{\includegraphics{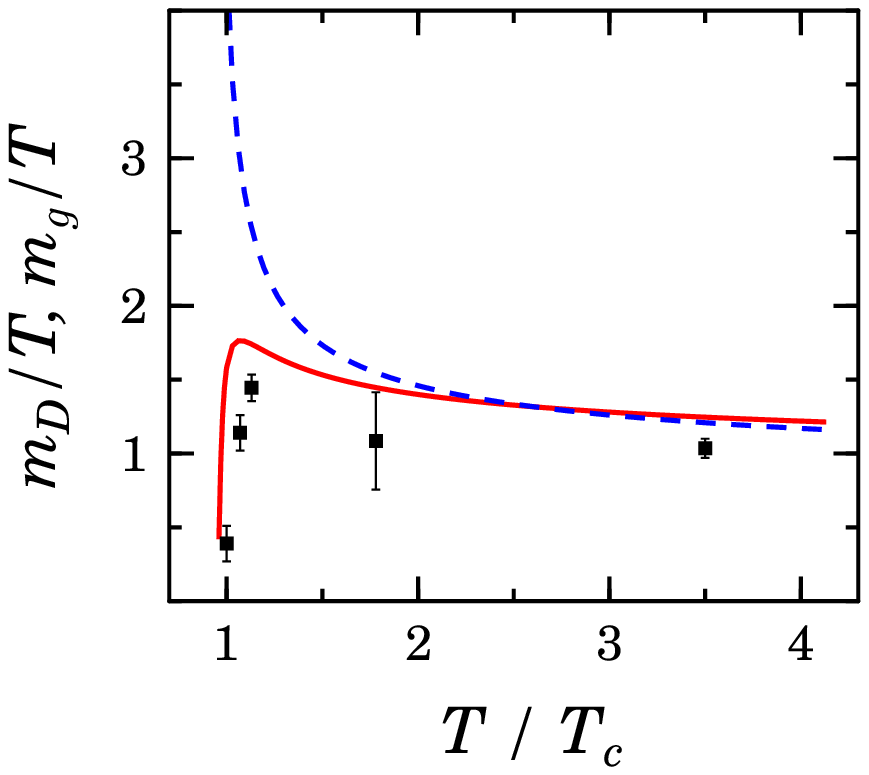}}
\end{flushleft}
\caption{\label{fig:quasiparticle} Extraction of $m_{\rm q}$ from 
  a fit of the lattice entropy, and comparison of the resulting
  $m_{\rm debye}$ with the Debye mass measured on the lattice (see the text for more details).}
\end{figure}

In practice, one could obtain the mass $m_{\rm q}$ from a
quasiparticle fit to the lattice entropy, as has been done in
\cite{PeshiKP1}. This is illustrated in figure
\ref{fig:quasiparticle}.
On the left plot, one can see that this fit reproduces perfectly the
entropy measured on the lattice, even for temperatures very close to
the critical temperature $T_c$. The quasiparticle mass (here it is a
gluon mass $m_g$, but this is proportional to $m_{\rm q}$) needed for
that fit is shown by a dashed line in the right plot, and one can see
that it becomes very large near $T_c$. The solid line is the Debye
mass calculated at 1-loop, with the mass obtained from the previous
fit used for the particle in the loop. The dots are the values of the
Debye mass measured on the lattice, and one can see that at least the
trend is remarkably well predicted by this very simple model. Since
photon rates often need to be evaluated at temperatures that are not
very large compared to $T_c$, it could be important to take the values
of $m_{\rm q}$ and $m_{\rm debye}$ from this model rather than from
the HTL approximation.

\section{LATTICE CALCULATIONS}
Recently appeared the first attempt to calculate directly on the
lattice the production rate of dileptons in a quark-gluon plasma. In
fact, the principle of this calculation has been known for a long
time: one should start from the Euclidean correlator of two vector
currents $\Pi(\tau,{\boldsymbol x})\equiv \left<j_\mu(0,{\boldsymbol
    0})j^\mu(\tau,{\boldsymbol x})\right>$, where $\tau\in[0,1/T]$ is
the Euclidean time. Next, one obtains $\Pi(\tau,{\boldsymbol q})$ by a
Fourier transformation of the spatial coordinates, and the imaginary
part of the real time self-energy is then related to this object by a
simple spectral representation:
\begin{equation}
{\Pi(\tau,{\boldsymbol q})}=\int_0^{\infty} \!\!{d\omega}\, 
{{\rm Im}\, \Pi_{\rm ret}{}^\mu{}_\mu(\omega,{\boldsymbol q})}\,
\frac{\cosh(\omega(\tau-1/2T))}{\sinh(\omega/2T)}\; .
\label{eq:lattice}
\end{equation}
In fact, this equation uniquely defines ${\rm Im}\, \Pi_{\rm
  ret}{}^\mu{}_\mu(\omega,{\boldsymbol q})$ if $\Pi(\tau,{\boldsymbol
  q})$ is known for all $\tau\in[0,1/T]$ and if one prescribes the
behavior of the solution at large $\omega$.

\begin{figure}[htbp]
\begin{flushleft}
\resizebox*{!}{4.5cm}{\rotatebox{-90}{\includegraphics{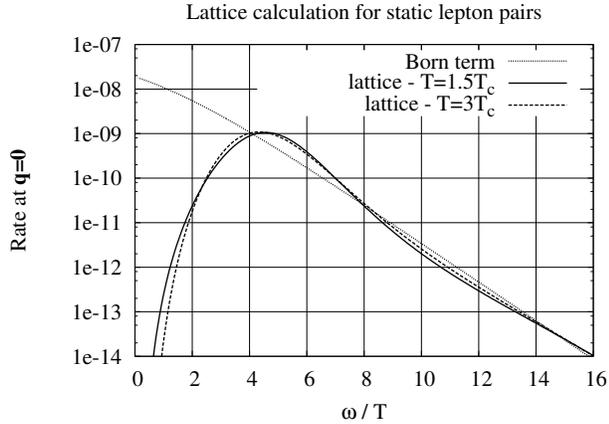}}}
\end{flushleft}
\caption{\label{fig:lattice} Lattice result for the production rate of 
static dileptons.}
\end{figure}

The main problem in lattice calculations is that one knows the
function $\Pi(\tau,{\boldsymbol q})$ only on the discrete points
of the lattice, which prevents from determining uniquely the solution.
This problem has been reconsidered recently using the Maximum Entropy
Method \cite{KarscLPSW1,Karsc1}, which is a way to take into account
prior knowledge about the solution (positivity, behavior at the
origin, etc...) in order to determine the most probable solution
compatible with the lattice data and with this a priori information.
The result obtained for static dileptons (${\boldsymbol q}=0$) via
this method is displayed in the figure \ref{fig:lattice}, for two
different values of the temperature. Note that this is a quenched
lattice simulation.
This result displays several interesting properties. At energies above
$4T$, the full rate is very close to the contribution of the Born
term, while at energies smaller than $3T$ it drops to extremely small
values. In addition, when plotted against $\omega/T$, the curves for
the two temperatures fall almost on top of one another, indicating
that the result scales like a universal function of $\omega/T$, at
least within the errors. 

In fact, the suppression at small $\omega$ has attracted a lot of
interest because it is not what one would expect from perturbation
theory: the resummation of thermal masses would indeed produce a drop
of the Born term because of threshold effects, but there are some
higher order processes that do not have a threshold and that should
fill the spectrum at small $\omega$. As of now, there are arguments
indicating that both the perturbative calculations and the lattice
calculation are incorrect at small $\omega$. If one evaluates
eq.~(\ref{eq:lattice}) at $\tau=1/2T$, one gets a sum rule:
$\int_0^\infty {d\omega} {{\rm Im}\,\Pi_{\rm
    ret}{}^\mu{}_\mu(\omega,{\boldsymbol q})}/{\sinh(\omega/2T)}=
\Pi(1/2T,{\boldsymbol q}) < \infty$, which is violated by all the
existing perturbative calculations (they give an infinite result)
because none of them includes the strong dissipative effects that
appear when one enters in the hydrodynamical regime ($\omega\to 0$).

On the other hand \cite{guy}, from the electric conductivity: $\sigma_{\rm
  el}=\lim_{\omega\to 0} {{{\rm Im}\,\Pi_{\rm
      ret}(\omega,{0})}/{6\omega}}$, one obtains ${\rm
  Im}\,\Pi_{\rm ret}(\omega,{0})\propto \omega$ when $\omega\to
0$. This implies that the static dilepton rate should diverge when
$\omega\to 0$. Unless the electric conductivity in quenched QCD is
nearly zero for some (yet to be explained) reason, the lattice
dilepton rate disagrees with this prediction at small $\omega$. Note
that `small' in these considerations means an $\omega$ small enough to
be in the hydrodynamical regime, i.e. $\omega \lesssim g_s^4 T$. In a
strong coupling theory, this regime could start as early as
$\omega\sim T$.

\section{CONCLUSIONS}
The full ${\cal O}(\alpha_s)$ photon and dilepton rates have now
been calculated using thermal QCD. This required to resum all the
diagrams involving multiple rescatterings as it turns out that the LPM
suppression plays a role at this order. A possible improvement of this
perturbative calculation could come via the use of quasiparticle
models adjusted to reproduce thermodynamical quantities determined on
the lattice, in order to get more realistic values for the various
mass parameters that describe the internals of the QGP.

Another recent development is the direct lattice evaluation of the
static dilepton rate. More work is still needed in this area in order
to fully understand the discrepancy with the perturbative approach at
low energy, and also to extend this program to real photons.

Finally, one should also mention that the way the pre-equilibrium
phase is treated (or, more accurately, ignored) is probably not
correct as there are indications that the kinetic equilibration time
might be as long as a few fermis \cite{Baier-eq}.  Determining how
quarks are dynamically generated from the initial gluons is a topic
that certainly deserves much more attention, and may influence
strongly the photon yields on obtains at the end.

\section*{Acknowledgement}

I would like to thank my collaborators P.~Aurenche, G.~Moore and
H.~Zaraket. I am also grateful to A.~Peshier and P.~Petreczky for
providing me some of the plots presented in this talk.

\bibliographystyle{unsrt} 

\end{document}